\documentclass[aps,showpacs,graphicx]{revtex4} 
\usepackage{amssymb} 
\usepackage[dvips]{graphicx} 
\usepackage[english]{babel} 
\usepackage{indentfirst} 
\usepackage{amsxtra} 
\usepackage{amsmath} 
\usepackage{supertabular} 
\usepackage{multirow} 
\usepackage [mathcal]{eucal} 
\usepackage {float}  
\usepackage{lscape} 
\usepackage{rotating}
\newcommand{\beq}{\begin{equation}} 
\newcommand{\eeq}{\end{equation}} 
\newcommand{\beqn}{\begin{eqnarray}} 
\newcommand{\eeqn}{\end{eqnarray}} 
\newcommand{\bsigma}{\mbox{\boldmath $\sigma$}} 
\newcommand{\btau}{\mbox{\boldmath $\tau$}} 
\newcommand{\half}{\frac{1}{2}} 

\newcommand{\br}{{\bf r}}

\newcommand{\fr}{{\cal F}}  % Depletion factor
\newcommand{\dr}{{\cal S}}  % flatness factor
\newcommand{\dev}{{\cal P}} % Deviation
\usepackage[usenames]{color} 
\usepackage{ulem}

\begin{document} 
 
\noindent 
\title{Matter distribution and spin-orbit force in spherical nuclei} 
 
\author{G. Co'$^{\,1,2}$, M. Anguiano$^{\,3}$,V. De Donno$^{\,1}$, 
A. M. Lallena$^{\,3}$ }
\affiliation{$^1$ Dipartimento di Matematica e Fisica ``E. De Giorgi'', 
  Universit\`a del Salento, I-73100 Lecce, ITALY, \\ 
$^2$ INFN Sezione di Lecce, Via Arnesano, I-73100 Lecce, ITALY, \\ 
$^3$ Departamento de F\'\i sica At\'omica, Molecular y 
  Nuclear, Universidad de Granada, E-18071 Granada, SPAIN
}  

\date{\today} 
 
%\bigskip 
 
\begin{abstract} 
We investigate the possibility that some nuclei show density distributions
with a depletion in the center, a semi-bubble structure, by 
using a Hartree-Fock plus Bardeen-Cooper-Schrieffer approach.
We separately study the proton, neutron and
matter distributions in 37 spherical nuclei mainly in the $s - d$ shell region. 
We found a relation between the semi-bubble structure and the 
energy splitting of spin-orbit partner single particle levels. The presence
of semi-bubble structure reduces this splitting, and we study its
consequences on the excitation spectrum of the nuclei under
investigation by using a quasi-particle random-phase-approximation approach.
The excitation energies of the low-lying $4^+$ states can be related
to the presence of semi-bubble structure in nuclei.
\end{abstract} 
 
%\bigskip 

 
\maketitle 

\section{Introduction} 
\label{sec:intro} 

The matter distribution of atomic nuclei is ruled 
by the interplay between the attraction of the
nucleon-nucleon interaction and the repulsion induced by the
Pauli exclusion principle and the Coulomb force. 
Since the short range of the interaction saturates the 
attraction effect, the global results is an almost constant matter distribution in the 
nuclear interior. This picture describes correctly the great majority of  
nuclei. However, there is a possibility that, in some cases, repulsive effects
dominate and, consequently, they produce a central depression in the matter
distribution that in the literature has taken the name of {\sl bubble} \cite{won72}. 
Following a commonly adopted nomenclature, we call
semi-bubble (SB) the nuclear systems with central depressions,
since with the term bubble is commonly called 
a distribution which is exactly zero at the center \cite{dec03}.

The possibility that some nuclei present a SB structure in 
the proton, neutron or matter distribution, is a problem 
which has been widely investigated by using various nuclear
models (see, for example
\cite{dav72,cam73,fri86,kha08,gra09,chu10,wan11,yao12,meu14,
li16,dug17,sch17}). 
In nuclei with a large number of protons, {\it i.~e.}
heavy and super-heavy nuclei, 
the SB features of the distributions are mainly due to
the Coulomb repulsion whose effects
become relevant \cite{sch17}. 
In this article, we address our attention
to medium-heavy nuclei in the $s-d$ shell region
where the source of eventual SB structures
is related to the Pauli  principle.

The best experimental tool to investigate 
matter distributions is the elastic electron 
scattering, even though it is mainly 
sensitive to the charge density \cite{bof96}.
These type of experiments can reach a sufficiently high resolution power
to allow the direct identification of SB structures.
Unfortunately, the nuclei where SB proton distributions
have been predicted are unstable, therefore they 
cannot be used as targets in traditional scattering
experiments. The facilities ELISe at FAIR \cite{sim07} 
and SCRIT at RIKEN \cite{sud09} 
are devised to carry out electron scattering experiments
on unstable nuclei. They use different
techniques and will become operative in the near
future. 

The technical difficulties outlined above have stimulated the search
for secondary, measurable, effects induced by, or directly related to,
the SB distributions. In all the mean-field (MF) 
descriptions of the nucleus, 
the effects on the total and single particle (s.p.) energies of the spin-orbit (s.o.)  
force are related to the derivatives of the matter, proton and neutron
distributions. The usual behavior of these 
distributions makes these derivatives to be almost zero in the nuclear interior, and negative
on the surface. The presence of a SB structure generates a positive derivative term
in the nuclear interior, and consequently a reduction of the effects of the s.o. force that 
could be observed by measuring the energy difference between 
s.o. partner levels in transfer reactions \cite{bur11t,bur14} or in sophisticated gamma ray 
detection experiments \cite{mut15t,mut17}.
This modification changes also the excitation spectrum. We have investigated wether the comparison
of spectra of isotopic or isotonic nuclei allows to identify the presence of a SB in proton or neutron densities. 

As already pointed out, in this article we investigate 
nuclei in the $s-d$ region of the nuclear chart 
to identify those which show a SB proton, neutron or
matter distribution, and the eventual consequences of this
feature. Since deformation can mask the effects induced 
by the SB structure, we have considered only spherical nuclei.

In our investigation we have used the Hartree-Fock (HF) 
plus Bardeen-Cooper-Schrieffer (BCS) approach
to describe the ground state of the nuclei we
have studied \cite{ang14}. In this way, 
pairing effects are taken into account 
in open shell nuclei. 
The excited states have been described 
by using a Quasi-particle Random Phase
Approximation (QRPA) approach \cite{don17}.
 Our calculations have been
carried out by consistently using the same effective nucleon-nucleon
interaction in each of the three steps, HF, BCS and QRPA.
We have used four different
finite-range interactions of Gogny type, with and without
tensor terms, in order to identify effects independent of the specific
input of the calculations. 

We present in Sec.~\ref{sec:details} the
features of our HF+BCS+QRPA  approach
interesting for the present study.
The following sections are dedicated to the presentation of
the results. In Sec.~\ref{sec:res:energ}
we compare the performances of our approach in the
description of the binding energies of the nuclei under 
investigation. In Sec.~\ref{sec:res:dens} we identify 
those nuclei showing SB structures. 
We investigate separately the proton, neutron and
matter distributions and we relate them to the 
shell structure generated by the four nucleon-nucleon
interactions we have considered. 
In Sec.~\ref{sec:res:so} we analyze the link between SB 
densities and the energy splitting between s.o. partner levels. 
In Sec.~\ref{sec:res:exc} we investigate the excitation
spectra of some SB nuclei to identify effects 
related to changes in the s.o. energy splitting.
A summary of our results is given in Sec.~\ref{sec:conclusions} 
where we also draw our conclusions. 

%%%%%%%%%%%%%%%%%%%%%%%%%%%%%%%%%%%
\section{The theoretical model}
\label{sec:details}
We describe the ground states
of the nuclei we have investigated by using a HF+BCS
approach. 
In  Refs.~\cite{ang14,ang15,ang16a}, we showed
that our HF+BCS calculations
produce results very closed to those obtained with 
the better grounded Hartree-Fock-Bogoliubov 
(HFB) theory. For the purposes of the present 
investigation, the differences between the results
of the two approaches are not relevant. 

The HF+BCS s.p. wave functions and the corresponding
occupation numbers have been used to 
describe the excited states within 
the QRPA theory presented in detail in Ref.~\cite{don17}.
In this reference we established the criteria 
for the numerical stability of all
the three steps of our calculations. 
In the present study we have adopted 
the same criteria also for those
nuclei which we investigate here for the first time. 

A crucial feature of our approach is the consistent use of the
same finite-range interaction  in each of the three steps 
of our calculations.  We have chosen to work with 
four different finite-range interactions. 
One of them is the traditional D1S Gogny force \cite{ber91}, widely
used in the literature, whose parameter values were chosen to reproduce 
the experimental values of a large set of binding energies and charge radii of 
nuclei belonging to various regions of the nuclear chart. 
However, this force has a well know
drawback: the neutron matter equation of state 
 has an unphysical behavior at large densities \cite{cha08}.
To solve this problem, the D1M parameterization was proposed \cite{cha07t,gor09}
This is the second interaction we have considered.

Together with these two parameterizations we 
have used the D1ST2a and the D1MTd forces, 
both containing tensor and tensor-isospin terms, and
constructed by following the strategy discussed in Refs. 
\cite{ang12, gra13,ang16a}. 
Starting with the original D1S and D1M parameterizations, respectively, 
we added two tensor terms of the form
\beq
V_{\rm T}(1,2) = \left[ {\cal V}_{\rm T1} + 
{\cal V}_{\rm T2} \btau(1) \cdot \btau(2)
\right] S_{12} \, \exp \left[-{(\br_1 - \br_2)^2 / \mu_{\rm T} }\right] \, ,
\label{eq:tensor}
\eeq
where we have indicated with $\btau$ the isospin of the nucleon, 
and with $S_{12}$ the traditional tensor operator \cite{bla52}.
A proper formulation of a new force would imply a global refit. 
However, since the observables used to choose
the values of the D1S and D1M parameters are essentially insensitive to 
the tensor force, we maintained  the original parameterizations 
of the central channels and selected the values of the parameters of the tensor force, 
${\cal V}_{\rm T1}$ and ${\cal V}_{\rm T2}$, 
in the following way: for the D1ST2a they were chosen
in order to reproduce the excitation energy of the $0^-$ state
in $^{16}$O and the energy splitting of the neutron $1f$ s.p. levels
in $^{48}$Ca (see Ref. \cite{ang12}), and for the D1MTd to properly
describe the excitation energies of the  $0^-$ states in $^{16}$O and in
$^{48}$Ca. These two nuclei are representative of the nuclear
chart regions we want to investigate, and the $0^-$ excited states and the splitting of spin orbit partners
are extremely sensitive to the tensor force \cite{ang11}. 
In both interactions the value of $\mu_{\rm T}$ has
been chosen to be equal to that of the Gaussian with 
the longest range in the D1S and D1M interactions, respectively (see Table~\ref{tab:tensor}).

\begin{table}[h] 
\begin{center} 
\begin{tabular}{lccc} 
\hline \hline
& ${\cal V}_{\rm T1}$ (MeV)  &  ${\cal V}_{\rm T2}$ (MeV) & $\mu_{\rm T}$ (fm) 
\\\hline
D1ST2a  & -135.0 & 115.0 & 1.2  \\ 
D1MTd  &  -230.0 & 180.0 & 1.0  \\ 
\hline \hline
\end{tabular}
\vskip -0.2 cm
\caption{\small Values of the parameters of the tensor force, defined in Eq.(\ref{eq:tensor}),\\ for the 
D1ST2a and D1MTd interactions. \label{tab:tensor} }
\end{center} 
\end{table} 

Comparing the results obtained with these four interactions, we have disentangled  
effects independent of the only arbitrary input of our approach
the effective nucleon-nucleon force. 
On the other hand, the main aim of our study is 
the relation between the presence of SB in the matter, 
proton and neutron distributions and the 
energy splitting of s.o. partner levels, which is rather sensitive
to $V_{\rm T}$ \cite{ots06}; therefore a 
comparison of results obtained with and without tensor terms 
in the interaction is mandatory.

In this study we have investigated 37 nuclei
having even $Z$ values between 8 and 26 and listed
in Table \ref{tab:benergy}.
All these nuclei
are spherical, according to the axially deformed HFB
calculations of Refs. \cite{cea,ang01a}, thus 
avoiding the possible complications that 
deformation would produce in the
identification of SB structures.

\begin{table}[t] 
\begin{center} 
\begin{tabular}{ccccccccccccccc} 
\hline \hline
element     &   $A$ &       D1M &      D1S & D1MTd & D1ST2a & exp &~~~& element & $A$ &      D1M &      D1S &  D1MTd & D1ST2a & exp \\ \hline
    O & $16$ & $-7.98$ & $-8.11$ & $-8.03$ & $-8.10$ & $-7.98$    & &  Ar & $38$ & $-8.52$ & $-8.63$ & $-8.58$ & $-8.63$ & $-8.61$  \\
       & $18$ & $-7.72$ & $-7.85$ & $-7.79$ & $-7.86$ & $-7.77$     & &      & $40$ & $-8.46$ & $-8.57$ & $-8.51$ & $-8.55$ & $-8.60$  \\\cline{9-15}
       & $20$ & $-7.54$ & $-7.66$ & $-7.62$ & $-7.68$ & $-7.57$     & & Ca & $34$ & $-7.15$ & $-7.26$ & $-7.21$ & $-7.26$ & $-7.20$ \\
       & $22$ & $-7.30$ & $-7.41$ & $-7.38$ & $-7.44$ & $-7.37$     & &      & $36$ & $-7.76$ & $-7.87$ & $-7.82$ & $-7.87$ & $-7.82$ \\
       & $24$ & $-6.94$ & $-7.04$ & $-7.03$ & $-7.08$ & $-7.04$     & &      & $38$ & $-8.18$ & $-8.30$ & $-8.24$ & $-8.30$ & $-8.24$ \\ \cline{1-7}
  Ne & $26$ & $-7.59$ & $-7.69$ & $-7.59$ & $-7.61$ & $-7.75$     & &      & $40$ & $-8.51$ & $-8.63$ & $-8.57$ & $-8.63$ & $-8.55$ \\
       & $28$ & $-7.24$ & $-7.34$ & $-7.27$ & $-7.30$ & $-7.39$     & &      & $42$ & $-8.56$ & $-8.67$ & $-8.62$ & $-8.67$ & $-8.62$ \\
       & $30$ & $-6.92$ & $-7.01$ & $-6.97$ & $-7.00$ & $-7.04$     & &      & $44$ & $-8.60$ & $-8.71$ & $-8.60$ & $-8.72$ & $-8.66$\\ \cline{1-7}  
 Mg & $28$ & $-8.05$ & $-8.17$ & $-7.98$ & $-8.00$ & $-8.27$     & &      & $46$ & $-8.61$ & $-8.72$ & $-8.69$ & $-8.73$ & $-8.67$ \\
       & $30$ & $-7.87$ & $-7.97$ & $-7.87$ & $-7.90$ & $-8.05$     & &      & $48$ & $-8.59$ & $-8.69$ & $-8.68$ & $-8.71$ & $-8.67$ \\
       & $32$ & $-7.69$ & $-7.78$ & $-7.75$ & $-7.77$ & $-7.80$     & &      & $50$ & $-8.46$ & $-8.55$ & $-8.54$ & $-8.57$ & $-8.55$ \\\cline{1-7} 
   Si & $30$ & $-8.31$ & $-8.45$ & $-8.21$ & $-8.23$ & $-8.52$     & &      & $52$ & $-8.32$ & $-8.40$ & $-8.34$ & $-8.43$ & $-8.43$ \\
       & $32$ & $-8.29$ & $-8.40$ & $-8.28$ & $-8.30$ & $-8.48$     & &      & $54$ & $-8.13$ & $-8.21$ & $-8.21$ & $-8.23$ & $-8.24$ \\
       & $34$ & $-8.25$ & $-8.33$ & $-8.31$ & $-8.33$ & $-8.34$     & &      & $56$ & $-7.94$ & $-8.01$ & $-8.00$ & $-8.01$ & $-8.04$ \\\cline{1-7} 
   S  & $30$ & $-7.92$ & $-8.06$ & $-7.84$ & $-7.87$ & $-8.11$     & &      & $58$ & $-7.76$ & $-7.81$ & $-7.81$ & $-7.81$ & $-7.84$ \\
       & $32$ & $-8.24$ & $-8.36$ & $-8.15$ & $-8.19$ & $-8.49$     & &      & $60$ & $-7.62$ & $-7.62$ & $-7.63$ & $-7.62$ &  --- \\ \cline{9-15}
       & $34$ & $-8.37$ & $-8.48$ & $-8.36$ & $-8.40$ & $-8.59$     & &  Ti & $42$ & $-8.19$ & $-8.31$ & $-8.25$ & $-8.31$ & $-8.26$ \\\cline{9-15}
       & $36$ & $-8.45$ & $-8.55$ & $-8.52$ & $-8.56$ & $-8.58$     & &  Cr & $44$ & $-7.89$ & $-8.01$ & $-7.96$ & $-8.02$ & $-7.96$ \\\cline{1-7}\cline{9-15}      
     &&&&&&                                                                                           & &  Fe & $46$ & $-7.57$ & $-7.69$ & $-7.65$ & $-7.71$ & $-7.62$ \\  %                                             
\hline \hline
\end{tabular}
\vskip -0.2 cm
\caption{\small  Binding energies per nucleon, in MeV, 
calculated in HF+BCS approach, for all the nuclei considered.
The experimental values are taken from \cite{bnlw}.
}
\label{tab:benergy}
\end{center} 
\end{table} 

%%%%%%%%%%%%%%%%%%%%%%%%%%
\section{Binding energies}
\label{sec:res:energ}
We list in Table \ref{tab:benergy} the binding energies per nucleon 
obtained with the four
interactions we have considered, and we compare them with
the experimental values taken from the compilation of
the Brookhaven National Laboratory \cite{bnlw}. 

\begin{table}[b] 
\begin{center} 
\begin{tabular}{rcccc}
\hline \hline
force &~~& $\overline{\Delta}(E_{\rm HF})$ &~& $\overline{\Delta}(E_{\rm HF+BCS})$ \\ \hline
D1M & &    0.021 (0.010) & & 0.013 (0.007) \\
D1MTd && 0.018 (0.017) & & 0.010 (0.012) \\ \hline
D1S &  &  0.011 (0.009) & & 0.007 (0.004)   \\
D1ST2a &&   0.017 (0.018) && 0.011 (0.010) \\
\hline \hline
\end{tabular}
\vskip -0.2 cm
\caption{\small Average values and standard deviations, in parentheses, 
of the relative differences with respect to the experimental 
binding energies, defined in Eq.~(\ref{eq:average}), 
obtained for the nuclei studied with the four interactions 
considered in the present work, in both HF and HF+BCS.
\label{tab:averages} }
\end{center} 
\end{table} 

To have a concise view of the agreement with 
the experimental data,  the relative differences 
\beq
\Delta(E_{a}) \,=\, \frac {|E_a \,-\, E_{\rm exp}| }{|E_{\rm exp}|}
\, ,
\label{eq:average}
\eeq
have been calculated for $a\equiv {\rm HF}$ and HF+BCS 
and for all the nuclei investigated.
In Table \ref{tab:averages} 
the average, $\overline{\Delta}(E_a)$, and the 
corresponding standard deviation are shown
for the four interactions considered. These results indicate the general good
agreement with the experimental values. In case of HF 
the average differences are about $2\%$ and the inclusion of BCS reduces them. 
The addition of the tensor terms to the interaction 
does not change sensitively the values obtained with D1M and D1S. 

\begin{figure}[h] 
\begin{center} 
\includegraphics [scale=0.4,angle=90]{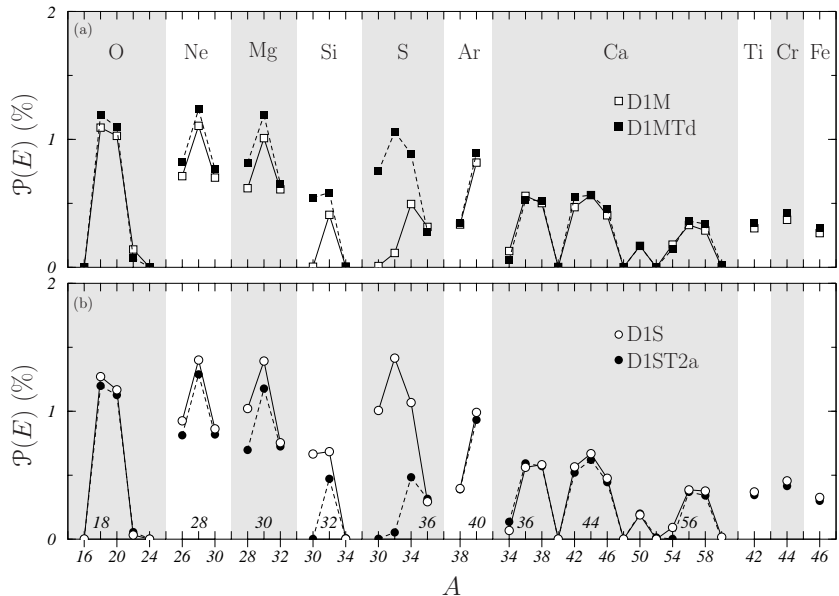} 
\vskip -0.3 cm 
\caption{\small Values of $\dev(E_{\rm HF+BCS, HF})$, 
defined in Eq. (\ref{eq:dev}), obtained with the four interactions
considered, for all the nuclei 
analyzed.
}
\label{fig:denergy} 
\end{center} 
\end{figure}

The values presented in Table \ref{tab:benergy} have been obtained
in HF+BCS calculations. An estimate of the effects of the pairing
is given in Fig. \ref{fig:denergy} where we show the 
so-called percentile deviations, defined as
\beq
\dev(E) \,= \, \frac {E_{\rm HF+BCS}\,-\, E_{\rm HF} }{E_{\rm HF+BCS} \,+\, E_{\rm HF}} 
\, ,
\label{eq:dev}
\eeq
for all the nuclei considered, and calculated
with the four interactions. In the figure, we do not observe 
remarkable differences between
the results obtained with the various forces. All the 
values are within 1.5\% indicating the small effect of the 
pairing on the binding energies of these
systems. The effect of the pairing on the nucleon density distributions
is discussed in the next section.

%%%%%%%%%%%%%%%%%%%%%%%%%%%%%%%%
\section{Density distributions}
\label{sec:res:dens}
A quantity widely used in the literature to
identify SB structures in density distributions is the 
{\sl depletion fraction}, which is defined as \cite{gra09}
\beq
\fr_\alpha = \frac{\rho_\alpha^{\rm max} - \rho_\alpha(0)}{\rho_\alpha^{\rm max}}
\, .
\label{eq:fr}
\eeq
Here $\rho_\alpha^{\rm max}$ is the maximum value reached by $\rho_\alpha(r)$, 
and $\alpha$ stands for proton (p), neutron (n) or matter (m).
The density distributions with SB structure have $\fr_\alpha > 0$. 

\begin{figure}[t] 
\begin{center} 
\includegraphics [scale=0.4,angle=90]{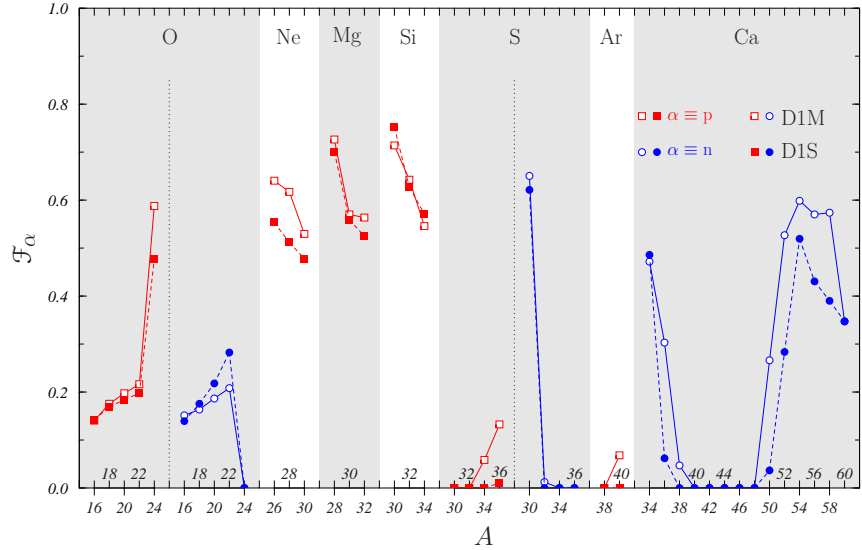} 
\vskip -0.3 cm 
\caption{\small Values of  the depletion fraction
$\fr$, defined in Eq. (\ref{eq:fr}),
for proton (red squares) and neutron (blue circles) density distributions. 
Open (solid) symbols indicate the results obtained with the
D1M (D1S) interaction. Lines are drawn to guide the eyes. }
\label{fig:eta} 
\end{center} 
\end{figure}

In Fig. \ref{fig:eta} we show the $\fr$ values 
obtained for proton (squares) and neutron (circles) 
density distributions for some of the nuclei investigated.
The values of $\fr_{\rm n}$ for the Ne, Mg, Si and Ar 
isotopes and those of $\fr_{\rm p}$ for the 
Ca nuclei are not shown because they are all zero. 
We found that the depletion fraction for the matter
distribution,  $\fr_{\rm m}$, is zero for all 
the nuclei studied except for the oxygen 
isotopes with $A<24$ where
$\fr_{\rm m}$ is of the same order of 
$\fr_{\rm p}$ and $\fr_{\rm n}$. In the 
figure, we compare the results 
obtained with the D1M (open symbols) and D1S 
(solid symbols) interactions. 
The two interactions containing 
the tensor terms produce $\fr$ values that are 
not sensitively different from 
those shown in the figure.

In general, in the nuclei having $\fr_{\rm p} > 0$, these are those showing a SB structure in the proton density,
the neutron depletion fraction, $\fr_{\rm n}$, is zero and vice-versa. There are, however, two 
exceptions to this trend. The first one is that of the oxygen 
isotopes from $A=16$ to 22 in which $\fr_{\rm p}$ and $\fr_{\rm n}$ are both, simultaneously,
positive. 
The second exception concerns the calcium isotopes from $A=40$ to 48 for which $\fr_{\rm p}=\fr_{\rm n}=0$. 

\begin{figure}[b] 
\begin{center} 
\includegraphics [scale=0.4,angle=90]{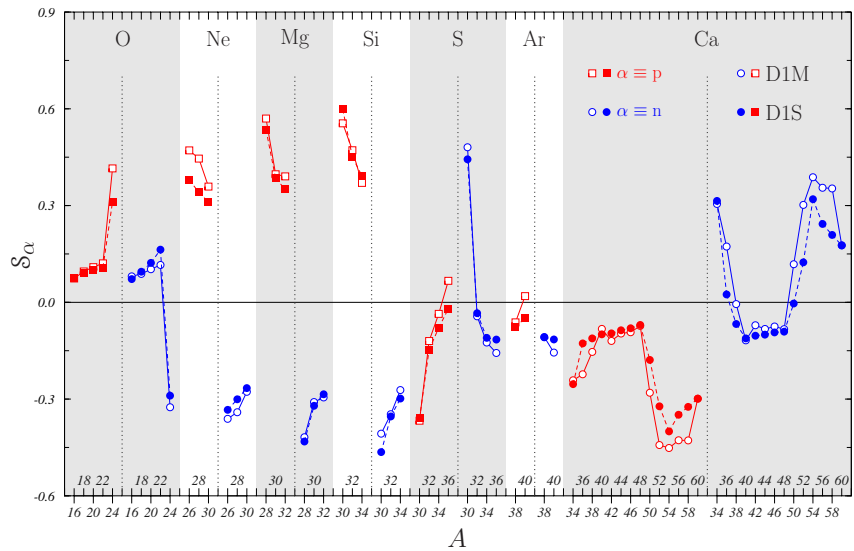} 
\vskip -0.3 cm 
\caption{\small Same as in Fig. \ref{fig:eta} for the flatness index
 $\dr$ defined in Eq. (\ref{eq:dr}). }
\label{fig:delta} 
\end{center} 
\end{figure} 

We investigate the presence of SB structures 
by using another quantity,
the \textsl{flatness index}, that we define as
\beq
\dr_\alpha = \frac{\rho_\alpha(r_{\rm mean}/2) - 
\rho_\alpha(0)}{\rho_\alpha(r_{\rm mean}/2) + \rho_\alpha(0)}
\, .
\label{eq:dr}
\eeq
In the above expression $r_{\rm mean}$ indicates the root mean  
squared radius of the density distribution and $\alpha$ 
has the same meaning as in Eq. (\ref{eq:fr}). An analogous
quantity has been used in \cite{sch17}.
Positive values of $\dr_\alpha$ indicate a SB structure in 
$\rho_\alpha$, and if $\dr_\alpha<0$ the corresponding 
density distribution 
has a maximum at the center of the nucleus. 
In general, the 
closer to zero is the value of $\dr_\alpha$ the 
flatter is the density
distribution in the nuclear interior.

The values of $\dr_{\rm p}$ and $\dr_{\rm n}$ for the nuclei 
investigated are
shown in Fig.~\ref{fig:delta}. As in the previous figure, 
squares (circles) indicate the results obtained for the 
proton (neutron) distributions, and open and solid symbols 
correspond to D1M and D1S interactions, respectively. 

In this figure, the trends already outlined
in discussing the $\fr_\alpha$ results become more evident: those nuclei having 
a $\rho_{\rm n}$ with a maximum at the nuclear center, for which $\dr_{\rm n}<0$, show
a SB proton density distribution ($\dr_{\rm p}<0$)
and vice-versa (with the exception of the oxygen 
isotopes above mentioned). We have found that the 
sum of the two densities, the matter distribution, 
is rather flat in all nuclei investigated, with
 $|\dr_{\rm m}|\sim 0.2$ at most.  

\begin{figure}[b]
\hspace*{0.5cm}
\begin{minipage}[c]{0.5\linewidth}
\includegraphics[width=\linewidth]{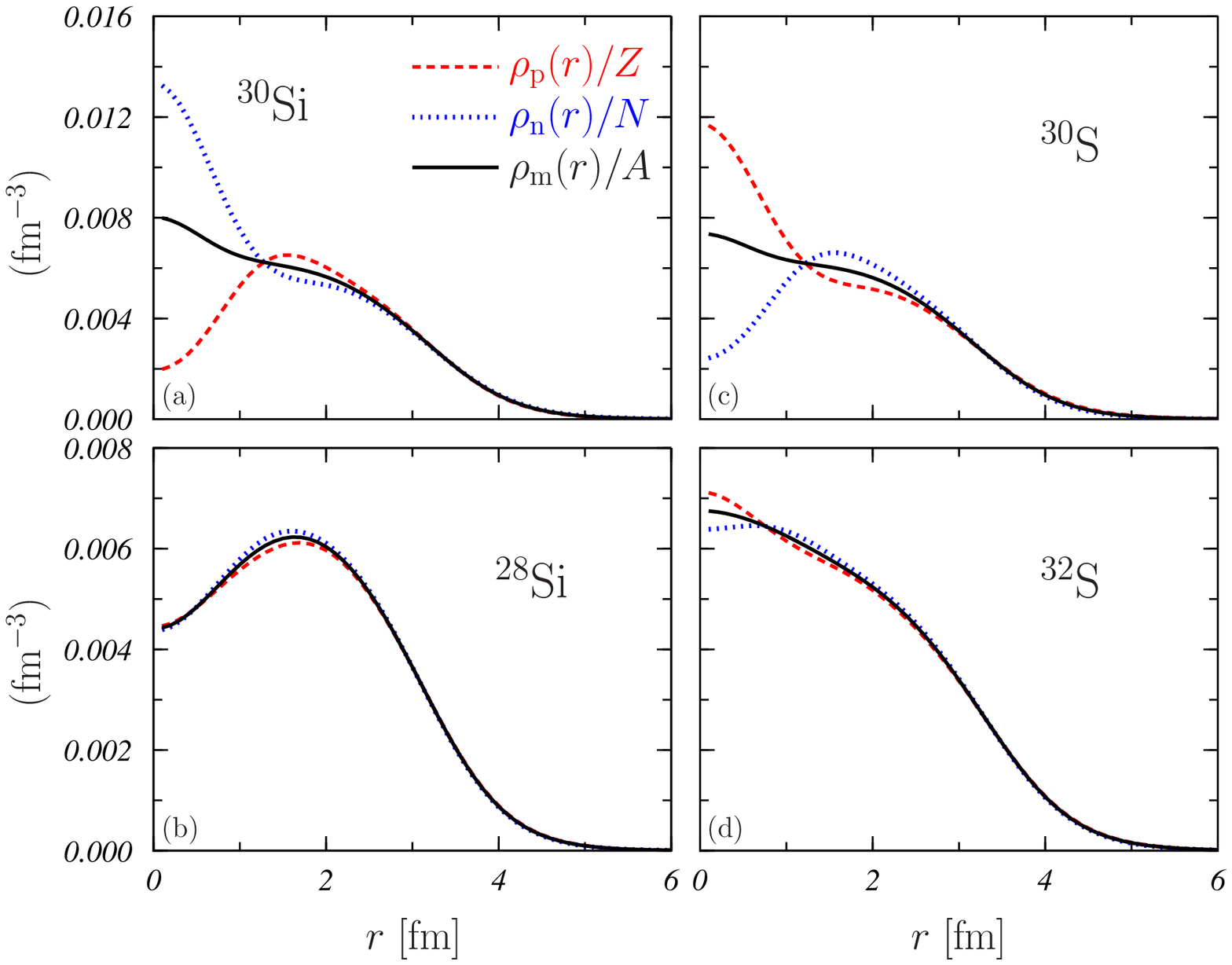}
\vskip -0.3 cm
\caption{\small Proton (red dashed lines), neutron (blue dotted lines)
and matter (black solid lines) density distributions, 
normalised to unity, for $^{28}$Si, $^{30}$Si, $^{30}$S and $^{32}$S. 
The calculations have been carried out with the D1M interaction. \label{fig:mgsi} }
\end{minipage}
\hspace*{1.cm}
\begin{minipage}[c]{0.4\linewidth}
\vspace*{-0.784cm} %\hspace*{-8cm}
\includegraphics[width=0.701\linewidth]{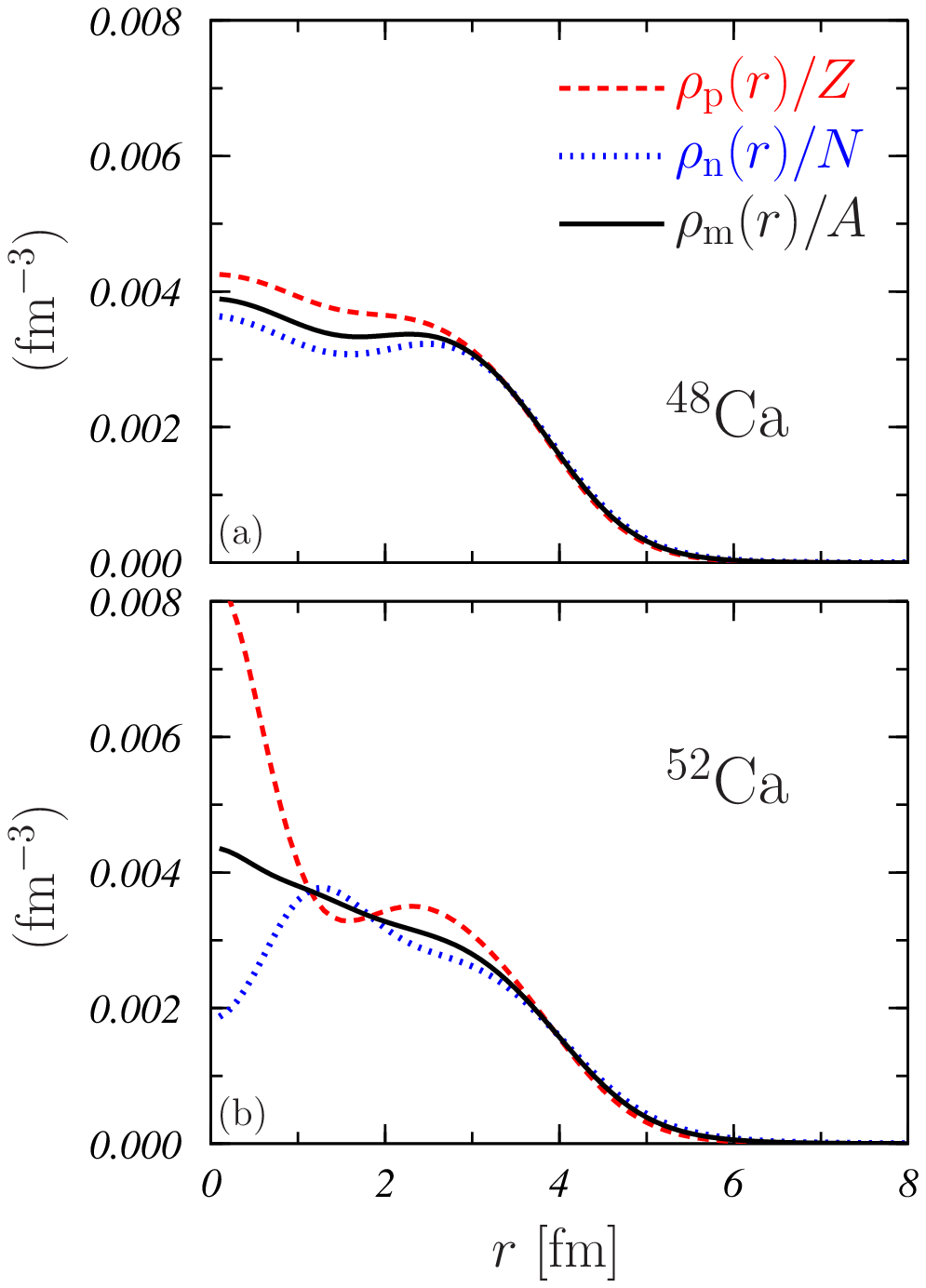}
\vskip -0.3 cm
\caption{\small Same as in Fig. \ref{fig:mgsi} but for the $^{48}$Ca \\and $^{52}$Ca nuclei. \label{fig:arca}}
\end{minipage}%
\end{figure}

A clear example of this compensation is provided by the
densities of the nuclei in the region around $A = 30$. 
In  Fig. \ref{fig:mgsi} we show the proton (red dashed curves), 
neutron (blue dotted curves) and matter  (black full curves) 
distributions of the $^{30}$Si, $^{30}$S, $^{28}$Si and 
$^{32}$S nuclei, calculated with the D1M interaction.
The densities obtained with the other three interactions 
are very similar.

The  two mirror nuclei $^{30}$Si and $^{30}$S have
high values of $\fr_{\rm p}$ and $\fr_{\rm n}$,
respectively. In $^{30}$Si, the $2s_{1/2}$ state 
is empty for protons and full for neutrons; 
as a consequence, $\rho_{\rm p}$ 
has a SB behaviour, while $\rho_{\rm n}$ 
has its maximum at the nuclear center. 
The opposite occurs in $^{30}$S. 
In both nuclei, the behaviours of
$\rho_{\rm p}$ and $\rho_{\rm n}$
counterbalance each other
producing matter distributions 
that do not show SB structures.

Another evidence that the occupancy of the 
$2s_{1/2}$ s.p. states is the source of the differences between the 
proton and neutron distributions can be visualized by looking at the
densities of $^{28}$Si (Fig.~\ref{fig:mgsi}b) and $^{32}$S (Fig.~\ref{fig:mgsi}d).
Since the $^{28}$Si nucleus is deformed, it
is not included in the set of the nuclei we have studied. 
It is, however, interesting to observe that
in our spherical HF+BCS 
description of this nucleus, both the proton and the neutron $2s_{1/2}$ 
levels are empty, and both $\rho_{\rm p}$ and $\rho_{\rm n}$ 
have SB structures, and, consequently, also  $\rho_{\rm m}$.
On the other hand, in $^{32}$S the neutron and proton 
$2s_{1/2}$ s.p. states have an occupation of 
95.7\% and 99.9\%, respectively. In this case $\fr_{\rm p}=0$ 
and $\fr_{\rm n}=0.012$ and the 
corresponding densities do not show SB structures. 
We can conclude that, in the region of nuclei with $A\sim30$,
the responsible of the appearance of SB
structures in proton or neutron density distributions is 
the occupancy of the $2s_{1/2}$ s.p. levels.

%\vspace*{23cm}
\begin{figure}[b] 
\begin{center} 
\includegraphics [scale=0.5,angle=0]{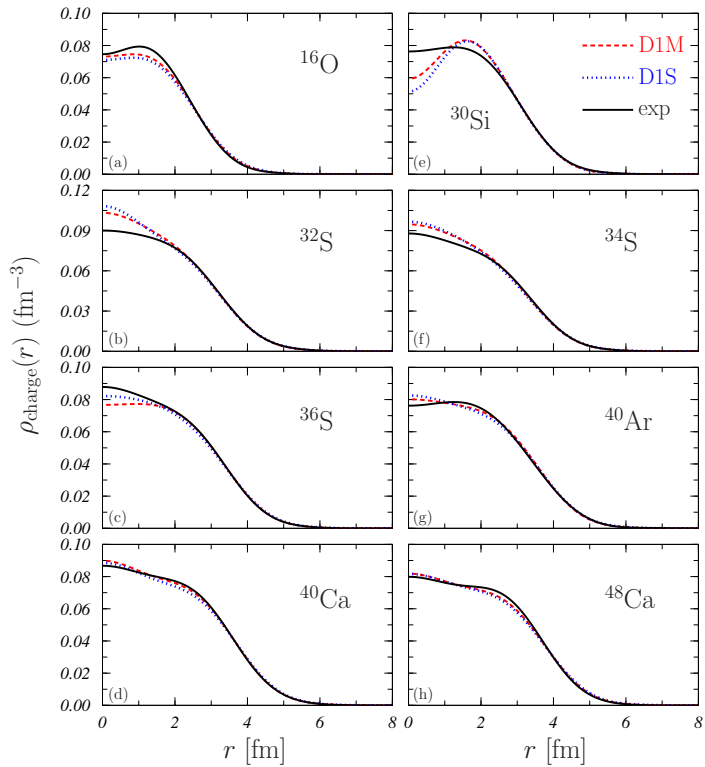} 
\vskip -0.3 cm
\caption{\small Charge distributions for some of the nuclei studied,
calculated with the D1M (red dashed curves) and D1S (blue dotted curves) interactions, 
compared with the empirical ones (black full curves) taken from the compilation 
of Ref. \cite{dej87}.
  }
\label{fig:ch6} 
\end{center} 
\end{figure} 

The situation is different in $A\sim50$ region. 
As example, we show in Fig. \ref{fig:arca} the density 
distributions of the $^{48}$Ca and $^{52}$Ca nuclei obtained 
with the D1M force. Similar results are found with the other interactions. 
In these two nuclei both the proton and neutron $2s_{1/2}$ 
s.p. levels are fully occupied. 
The proton densities in both nuclei have a maximum at $r=0$, and also
$\rho_{\rm n}$ in $^{48}$Ca, while in $^{52}$Ca  the neutron distribution
show a SB structure.
This behavior is due to the filling of the neutron 
$2p_{3/2}$ s.p. level in $^{52}$Ca. The contribution of
this state is very small at $r\sim0$ but 
it is remarkable at $r\sim1.5-2\,$fm.  
In this nucleus, as well as in those in the
same mass region,
the appearance	of a SB structure is due
to the filling of s.p. states peaked slightly far from the nuclear center.

\begin{figure}[t] 
\begin{center} 
\includegraphics [scale=0.45,angle=0]{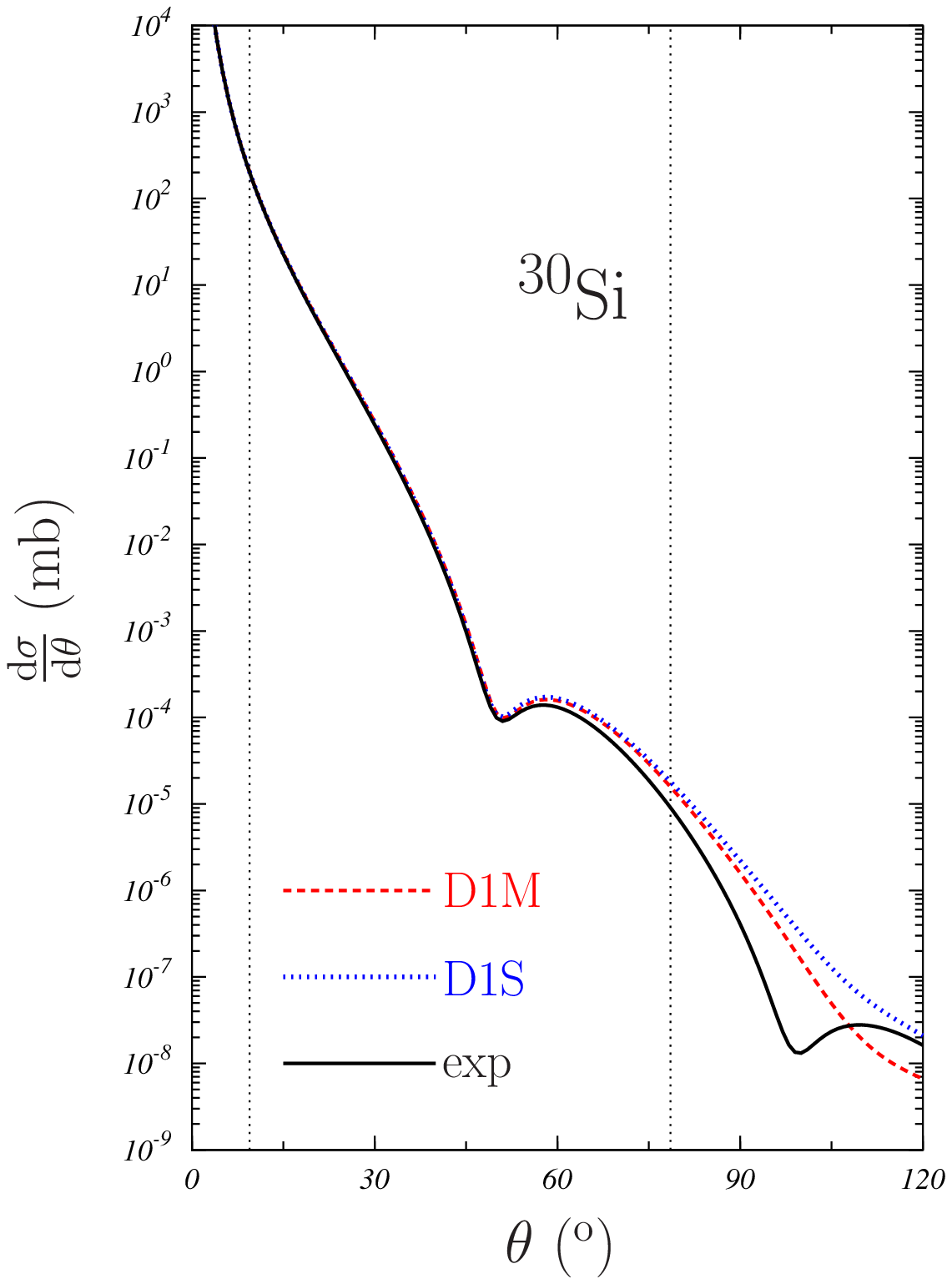} 
\vskip -0.3 cm
\caption{\small Elastic electron scattering cross sections on $^{30}$Si
target calculated for an incident electron energy of $300\,$MeV, 
within the distorted wave Born approximation
by using the charge densities shown in panel (b) of Fig. \ref{fig:ch6}. 
The results obtained with the D1M (red dashed curve) and 
D1S (blue dotted curve) interactions are compared to those 
corresponding to the empirical density (black solid curve). 
The two vertical lines indicate the range probed by the 
experiment as indicated in Ref. \cite{dej87}. }
\label{fig:xssi} 
\end{center} 
\end{figure} 

In Fig. \ref{fig:ch6} we compare the available empirical
charge density distributions taken from the compilation
of Ref. \cite{dej87} with those we have obtained  
by folding the proton densities
with the traditional proton
dipole electromagnetic form factor
\cite{pov93}. The use of more accurate 
form factors \cite{hoe76} produces differences 
within the numerical accuracy of our 
calculations. The results obtained with the D1S and D1M 
interactions are rather similar, especially on 
the surfaces of the nuclei. The main differences
between the various distributions
are localized in the nuclear interior.

The empirical charge distributions 
are obtained by
fitting elastic electron scattering data that 
cover a given range of momentum
transfer, $q$, values. By considering 
that the resolution power is inversely 
proportional to the maximum momentum transfer involved, the more
accurate experiments are those done on $^{16}$O, $^{40}$Ca
and $^{48}$Ca nuclei where $q_{\rm max} \simeq 3 - 3.7 \, {\rm fm}^{-1}$.
The sulfur, $^{30}$Si and $^{40}$Ar 
data have been taken with $q$ up to $2.6$,
1.5 and $1.8\,$fm$^{-1}$, respectively. 
It is possible that the largest differences observed 
between our charge densities and the experimental 
ones, that occur for $^{30}$Si (panel (e)) and 
$^{32}$S (panel (b)) can be fictitious because of the limited 
resolution power obtained by the experiments.

We illustrate this point by considering the 
example of the $^{30}$Si nucleus. In this case, 
our calculations predict
a remarkable SB structure for both interactions,
while the experiment indicates an almost flat density.
By using the charge distributions shown in that figure,
we have calculated elastic electron scattering cross 
sections in distorted wave Born approximation. 
In Fig. \ref{fig:xssi} we show the results obtained for 
electrons with an incident energy of $300\,$MeV. 
The vertical lines indicate the angles corresponding 
to the smallest and largest momentum transfer  
probed by the experiment and used to extract the empirical
density \cite{dej87}. 
We observe that the differences between
empirical and theoretical densities shown in the 
panel (b) of Fig.~\ref{fig:ch6} generate 
noticeable effects only at large scattering angles, 
in a region outside the $q$ range probed by the experiment, 
indicate by the two vertical lines. 

The results of Fig.~\ref{fig:ch6} show that, for all the nuclei considered,
the empirical charge densities are well described by our results at the nuclear surface.
This is confirmed in Table \ref{tab:rms}, where the 
available experimental charge rms radii, 
taken from the compilation of Ref.~\cite{ang13},
are compared to those we have obtained with 
the four interactions considered.
The maximum relative deviation from the experimental data is 
around $3\%$. 
This good agreement is not surprising since charge 
rms radii were inserted in the fit procedure to determine the 
values of the parameters of the D1S and D1M interactions \cite{cha07t}. 

\begin{table}[!b] 
\begin{center} 
\begin{tabular}{ccccccccccccccc} 
\hline \hline
element     &   $A$ &       D1M &      D1S & D1MTd & D1ST2a & exp &~~~& element & $A$ &      D1M &      D1S &  D1MTd & D1ST2a & exp \\  \hline
 O  & 16 & $ 2.76$ & $ 2.79$ & $ 2.76$ & $ 2.79$ & $ 2.70$ & & Ar  & 38 & $ 3.39$ & $ 3.42$ & $ 3.39$ & $ 3.43$ & $ 3.40$  \\
      & 18 & $ 2.77$ & $ 2.80$ & $ 2.77$ & $ 2.80$ & $ 2.77$ & &      & 40 & $ 3.39$ & $ 3.43$ & $ 3.39$ & $ 3.43$ & $ 3.43$ \\  \cline{9-15}
      & 20 & $ 2.78$ & $ 2.81$ & $ 2.78$ & $ 2.81$ & --- & &           Ca & 34 & $ 3.47$ & $ 3.51$ & $ 3.46$ & $ 3.50$ & --- \\
      & 22 & $ 2.80$ & $ 2.82$ & $ 2.79$ & $ 2.82$ & --- & &                & 36 & $ 3.47$ & $ 3.49$ & $ 3.46$ & $ 3.49$ & --- \\
      & 24 & $ 2.80$ & $ 2.83$ & $ 2.79$ & $ 2.83$ & --- & &                & 38 & $ 3.46$ & $ 3.49$ & $ 3.46$ & $ 3.49$ & ---\\ \cline{1-7}
Ne & 26 & $ 2.91$ & $ 2.95$ & $ 2.92$ & $ 2.97$ & $ 2.93$ & &       & 40 & $ 3.46$ & $ 3.50$ & $ 3.46$ & $ 3.50$ & $ 3.48$ \\
      & 28 & $ 2.97$ & $ 3.01$ & $ 2.97$ & $ 3.02$ & $ 2.96$ & &       & 42 & $ 3.48$ & $ 3.51$ & $ 3.47$ & $ 3.51$ & $ 3.51$ \\
      & 30 & $ 3.03$ & $ 3.07$ & $ 3.03$ & $ 3.07$ & --- & &                 & 44 & $ 3.49$ & $ 3.52$ & $ 3.49$ & $ 3.52$ & $ 3.52$ \\ \cline{1-7}
Mg & 28 & $ 2.99$ & $ 3.02$ & $ 3.00$ & $ 3.06$ & ---                & &  & 46 & $ 3.50$ & $ 3.52$ & $ 3.49$ & $ 3.52$ & $ 3.50$ \\
      & 30 & $ 3.06$ & $ 3.10$ & $ 3.07$ & $ 3.11$ & ---                & &  & 48 & $ 3.50$ & $ 3.53$ & $ 3.50$ & $ 3.53$ & $ 3.48$ \\
      & 32 & $ 3.11$ & $ 3.15$ & $ 3.11$ & $ 3.16$ & ---                & &  & 50 & $ 3.56$ & $ 3.57$ & $ 3.56$ & $ 3.57$ & $ 3.52$\\\cline{1-7} 
Si   & 30 & $ 3.06$ & $ 3.08$ & $ 3.08$ & $ 3.12$ & $ 3.13$     & &  & 52 & $ 3.62$ & $ 3.62$ & $ 3.67$ & $ 3.62$ & ---  \\
      & 32 & $ 3.12$ & $ 3.15$ & $ 3.13$ & $ 3.17$ & ---                & &  & 54 & $ 3.72$ & $ 3.72$ & $ 3.71$ & $ 3.72$ & --- \\
      & 34 & $ 3.18$ & $ 3.21$ & $ 3.18$ & $ 3.21$ & ---                & &  & 56 & $ 3.72$ & $ 3.73$ & $ 3.72$ & $ 3.73$ & ---\\\cline{1-7} 
S   & 30 & $ 3.25$ & $ 3.26$ & $ 3.26$ & $ 3.29$ & ---                & &  & 58 & $ 3.74$ & $ 3.75$ & $ 3.74$ & $ 3.75$ & ---  \\
     & 32 & $ 3.24$ & $ 3.27$ & $ 3.26$ & $ 3.30$ & $ 3.26$      & &  & 60 & $ 3.77$ & $ 3.77$ & $ 3.75$ & $ 3.77$ & --- \\ \cline{9-15}
     & 34 & $ 3.27$ & $ 3.29$ & $ 3.28$ & $ 3.31$ & $ 3.29$   & & Ti & 42 & $ 3.55$ & $ 3.59$ & $ 3.55$ & $ 3.59$ & --- \\ \cline{9-15}
     & 36 & $ 3.29$ & $ 3.33$ & $ 3.29$ & $ 3.33$ & $ 3.30$  & & Cr & 44 & $ 3.62$ & $ 3.66$ & $ 3.61$ & $ 3.66$ & --- \\ \cline{1-7} \cline{9-15}
    &&&&&&                                                                             & & Fe & 46 & $ 3.67$ & $ 3.72$ & $ 3.67$ & $ 3.72$ & --- \\
\hline \hline
\end{tabular}
\vskip -0.2 cm
\caption{\small  Charge root mean square radii, in fm, calculated in HF+BCS approach, for all the nuclei considered, with the four interactions used in this work. The experimental values are taken from Ref.~\cite{ang13}.
\label{tab:rms} }
\end{center} 
\end{table} 

We conclude this section briefly discussing the role of the pairing
on the density distributions. In general, the effect of the pairing 
on this observable is negligible. 
There are, however, some remarkable exceptions to this general trend. 
The pairing reduces the $\fr$ values for the neutron distributions of $^{20}$O 
and $^{22}$O of about the 10\%. 
We found larger effects on the $\fr$ values of the 
proton distributions of $^{36}$S and $^{40}$Ar and 
those of the neutron 
distributions of $^{36}$Ca and $^{38}$Ca which are almost doubled
by the inclusion of the pairing.

%%%%%%%%%%%%%%%%%%%%
\section{Spin-orbit splitting}
\label{sec:res:so}

We have pointed out in the previous section 
that the experimental investigation of the presence 
of a SB structure in the proton, or charge, density 
distribution requires a high spatial resolution, therefore 
elastic electron scattering experiments involving high values 
of the momentum transfer.  However, these experiments are
rather difficult to carry out, especially on unstable nuclei. 
In the following we address the 
question whether there are subsidiary observables
that can be related to the occurrence of SB structures in nuclear densities.

According to Ref. \cite{vau72}, the contribution of the
s.o. term of the interaction to the total energy of the system is given by
\beq
E_{\rm s.o.} \,=\, \half \, W_0 
\int {\rm d}^3 r \, (\nabla \rho_{\rm m} \cdot {\bf J_{\rm m}} \,+\, \nabla \rho_{\rm n} \cdot {\bf J}_{\rm n} 
\,+\, \nabla \rho_{\rm p} \cdot {\bf J}_{\rm p} )
\, ,
\label{eq:eso}
\eeq
where the spin density ${\bf J}_\tau$ is defined as
\beq
{\bf J}_\tau(\br) \,=\, -\,i\, \sum_{\alpha \sigma \sigma'} [\phi^ \tau_{\alpha \sigma} (\br)]^* \,
\left[ \nabla \phi^\tau_{\alpha\sigma'}(\br) \times \langle \sigma | \bsigma | \sigma' \rangle
\right] \,,Ê\,\,\,\,\, \tau \equiv {\rm p},{\rm n}
\, ,
\eeq
and ${\bf J}_{\rm m}={\bf J}_{\rm p}+{\bf J}_{\rm n}$.
In the above equations
$\phi$ indicates the s.p. wave function
characterized by the third components of the isospin, $\tau$, and of the spin, 
$\sigma$, and by other quantum numbers $\alpha$. 

As we can see in Eq. (\ref{eq:eso}), the value of
$E_{\rm s.o.}$ depends on the derivative of the density 
distributions. Therefore, the presence of
a SB structure could show up in the s.o. energy because 
it would make $\nabla \rho$ to behave in the nuclear interior 
with opposite sign with respect to the
surface, thus producing an overall reduction. 
We investigate if some observable linked to the s.o. 
interaction can be used to reveal SB structure
in nuclei. The relationship between the s.o. interaction and the SB structure in
density distribution has been investigated in Refs. 
\cite{gra09,wan11,li16,dug17,bur11t,bur14,mut15t,shu16,mut17,kar17,tod04,wu14}.

We have estimated the order of magnitude of the effect described above by
performing a toy calculation for the $^{40}$Ca nucleus. First, we have obtained 
two sets of s.p. wave functions from a mean-field potential of the form
\beq
V(r) \,=\, \frac {V_0} {1\,+\,\exp \left( \displaystyle \frac{r\,-\,R}{a} \right) } \,+ \, 
B \, \exp  \left( - \displaystyle \frac{r^2}{b^2} \right) \, .
\label{eq:woods}
\eeq
The second term of the above equation has been used to generate
density profiles with SB structure.

The values $V_0=-50\,$MeV, $R = 4.4\,$fm, $a=0.6\,$fm, and $b= 0.5\,$fm 
have been chosen for both protons and neutrons. 
We have considered $B=0$ and $B=140\,$MeV 
to obtain the proton s.p. wave function of the two sets; for the neutron ones we used $B=0$ 
in the two cases. Using these s.p. wave functions, the direct Hartree, $\Gamma_{\rm d}(\br)$,
and the exchange Fock-Dirac, $\Gamma_{\rm ex}(\br,\br')$, potentials \cite{rin80} entering into
the HF equations have been calculated for the D1M interaction and then we have solved the
one-body Schr\"odinger equation 
\beq
-\frac{\hbar^2}{2m} \nabla^2 \phi_k(\br) \,
+ \,  \Gamma_{\rm d}(\br)\, \phi_k(\br)\,
- \, \int {\rm d}\br' \, \Gamma_{\rm ex}(\br,\br')\, \phi_k(\br') 
\,=\, \epsilon_k \, \phi_k(\br) 
\label{eq:toy}
\eeq
to obtain the corresponding s.p. states as well as their energies.  

\begin{figure}[!b] 
\begin{center} 
\includegraphics [scale=0.5,angle=0]{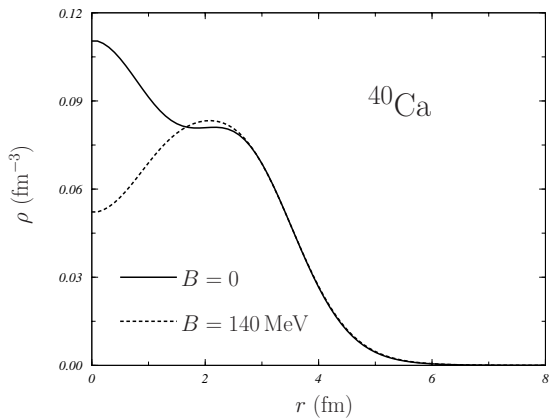} 
\vskip -0.3 cm 
\caption{\small Proton distribution for the $^{40}$Ca nucleus obtained by using the
mean-field potential of Eq. (\ref{eq:woods}). The full line has been obtained without
the gaussian term, $B=0$, in Eq. (\ref{eq:woods}). The dashed line is the
result obtained by inserting the gaussian term with the parameters indicated in the text.
  }
\label{fig:woods} 
\end{center} 
\end{figure}

We show in Fig. \ref{fig:woods}
the proton density distributions 
corresponding to the full potential (dashed curve) 
and to the potential with $B=0$ (solid curve). 
The former has a SB structure with $\fr_{\rm p}
=0.37$. From the solutions of Eq.~(\ref{eq:toy})
we have evaluated the s.o. splitting 
\beq
s^\alpha_{nl} \,=\, \epsilon_{nlj-1/2} \,-\, \epsilon_{nlj+1/2} \, , \,\,\,\, \alpha \equiv {\rm p,n} \, ,
\label{eq:sosplit}
\eeq
where $\epsilon$ labels the s.p. energy and $n$, $l$ and $j$ are the 
quantum numbers characterizing the 
state. In the calculation with $B=0$, {\it i.e.} without SB structure, 
we have found that $s^{\rm p}_{1p}=3.24\,$MeV 
and $s^{\rm p}_{1d}=4.84\,$MeV. 
These values reduce to $0.53\,$MeV and $3.40\,
$MeV when the densities with SB structure are used.

\begin{figure}[b] 
\begin{center} 
\includegraphics [scale=0.4,angle=90]{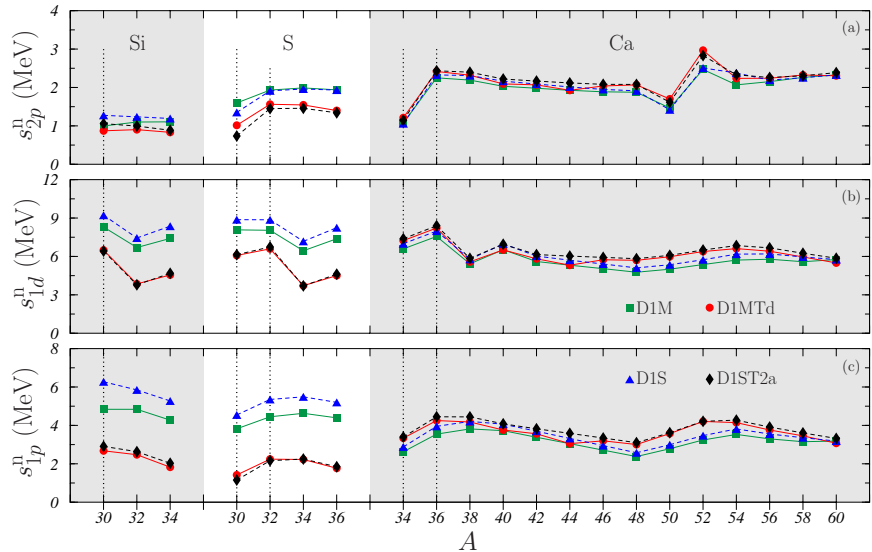} 
\vskip -0.3 cm 
\caption{\small HF+BCS spin orbit splitting $s_{nl}$, defined in Eq. (\ref{eq:sosplit}),
for neutron $1p$, $1d$ and $2p$ s.p. states. The results obtained for the Si, S, Ar and Ca isotope 
chains are shown as a function of $A$. Green squares, 
blue triangles, red circles and black diamonds indicate the 
values calculated with the D1M, D1S, D1MTd and D1ST2a 
interactions, respectively.
The vertical dotted lines indicate the nuclei with $N=14$ and $N=16$. }
\label{fig:splitz} 
\end{center} 
\end{figure}

Once the order of magnitude of the expected effect, 
given by the reduction of the $s_{nl}$ values just discussed, 
was evaluated, we analysed the s.o. splittings obtained in 
our HF+BCS calculations for all the nuclei we are 
studying. In Fig. \ref{fig:splitz} we show the neutron splittings $s^{\rm n}_{2p}$ 
(panel (a)), $s^{\rm n}_{1d}$ (panel (b)) and $s^{\rm n}_{1p}$ 
(panel (c)) as a function of $A$ for the Si, S, 
and Ca isotope chains. Green squares, 
blue triangles, red circles and black diamonds 
indicate the values calculated with the D1M, D1S, D1MTd and D1ST2a 
interactions, respectively. 

The only recurring effect we observe 
is an increase of the splitting 
when going from $N=14$ to $N=16$; these values of $N$ are
represented by vertical dotted lines in the figure. 
This behavior is related to the filling of the neutron 
$2s_{1/2}$ level. In the case of $N=14$ this level is empty and this
generates a SB neutron distribution, as can be checked in 
Figs. \ref{fig:eta} and \ref{fig:delta}. All the nuclei 
with $N=16$ have the neutron $2s_{1/2}$ level fully occupied 
and the corresponding neutron densities do 
not show SB structure.

A similar increase in the splitting is observed between $N=30$ and 
$N=32$ in  Ca, for the $2p$ level only, and between $N=18$ 
and $N=20$ for the $1d$ state in the three isotope chains. 
In these cases, the effect is linked to the fact that the neutron
 $2p_{3/2}$ and $1d_{3/2}$ states are fully occupied
 for $N=32$ and $N=20$, respectively. However, while in 
 the first case, the  Ca involved present a SB structure 
 in the neutron density, with a $\dr_{\rm n}$ value that 
 increases with the splitting, none of the nuclei with $N=18$ 
 and $N=20$ show SB neutron densities 
 (see Figs. \ref{fig:eta} and \ref{fig:delta}). 

\begin{figure}[t] 
\begin{center} 
\includegraphics [scale=0.4,angle=90]{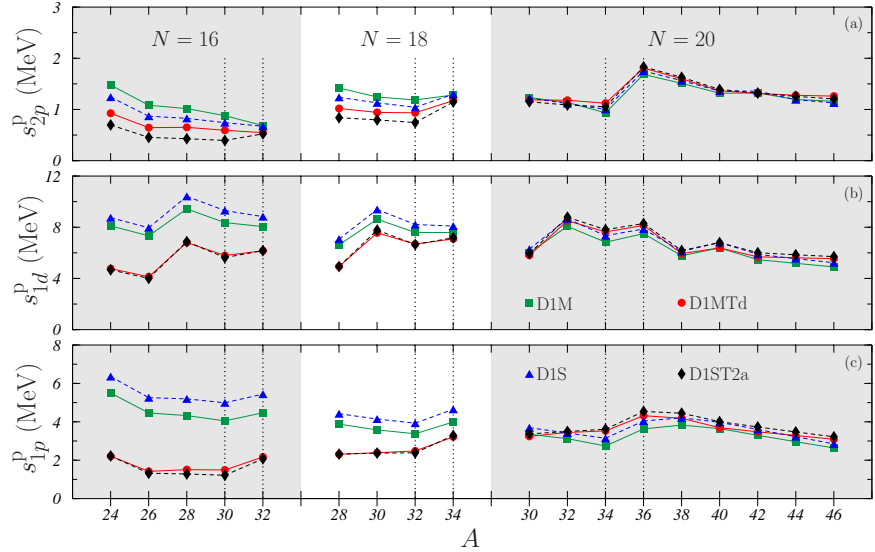} 
\vskip -0.3 cm 
\caption{\small HF+BCS spin orbit splitting $s_{nl}$, as defined in Eq. (\ref{eq:sosplit}),
for proton $1p$, $1d$ and $2p$ s.p. states. The results obtained for the $N=16$, 18, 20 and 22 isotone 
chains are shown as a function of $A$. Green squares, 
blue triangles, red circles and black diamonds indicate the 
values calculated with the D1M, D1S, D1MTd and D1ST2a 
interactions, respectively.
The vertical dotted lines indicate the nuclei with $Z=14$ and $Z=16$. 
 }
\label{fig:splitn} 
\end{center} 
\end{figure}

We observe an analogous behaviour in Fig. \ref{fig:splitn} where 
we show, as a function of $A$, the values of $s^{\rm p}_{nl}$ 
for the $2p$ (panel (a)), $1d$ (panel (b)) and $1p$ (panel (c)) 
proton s.p. states and for the isotone chains with $N=16$, 18, and 
20. Again, $s_{nl}$ increases between $Z=14$ and $Z=16$ 
(indicated with the dotted vertical lines in the figure), the nuclei 
with $Z=14$ having SB proton densities, while those with 
$Z=16$ do not. This behaviour is related to the occupation 
of the proton $2s_{1/2}$ s.p. level. However, some exceptions 
occur in $N=16$: $s^{\rm p}_{2p}$, for D1M, D1S and 
D1MTd, and $s^{\rm p}_{1d}$, for D1M and D1S.
As in the case of Fig. \ref{fig:splitz}, we observe
a  systematic increase of the $s_{1d}$ values
from $Z=18$ to $Z=20$ in all the interactions for $N=20$.

Both Figs. \ref{fig:splitz} and \ref{fig:splitn} show that 
the increase of the $s_{nl}$ values when 
the number of proton or neutrons changes from 14 to 
16 occurs for the four interactions considered,
independently of the inclusion of tensor terms. 
This is important since the s.o. 
splittings may be strongly influenced by these terms 
\cite{ots06}, as it can be seen for $1d$ and $1p$ s.p. levels in Si and S 
(Fig. \ref{fig:splitz}) and in $N=16$ (Fig. \ref{fig:splitn}). 
In these cases the results obtained with the 
interactions containing tensor terms are about 2 MeV
smaller than those obtained with the D1M and D1S forces.
This is a consequence
of the effect described by Otsuka 
\cite{ots06} that predicts a reduction of the splitting 
between the energies of the s.o. partner levels
due to the contribution of the unlike particle term of the tensor force. 
This effect becomes smaller in nuclei with proton or neutron s.o. saturated 
levels.

In Refs. \cite{bur11t, mut15t} it is reported that the s.o. 
splitting $s_{2p}^{\rm p}$ of $^{36}$S is larger than 
that  of $^{34}$Si. This is in agreement 
with the results of our calculations.

\section{Excitation spectra}
\label{sec:res:exc}

The instability of the nuclei where we have identified the 
presence of SB structures generates an intrinsic difficulty in 
using them as targets of  scattering experiments. On the other 
hand, the study of the excitation spectra is a,
relatively, easier task. For this reason, after having verified 
that a SB density distribution is linked to the size of
the splitting between the energies of the s.o. partner 
levels, we analyzed how this 
influences the excitation spectra. 

\begin{table}[!b] 
\begin{center} 
\begin{tabular}{c c c c c c c} 
\hline \hline
 $J^\pi$ & nucleus & D1M & D1S & D1MTd & D1ST2a & exp  \\
\hline
  $2^+$ &$^{34}$Si & 3.99 & 4.30 & 4.00 & 4.25 & 3.33 \\
   &  $^{36}$S & 1.81 & 2.03 & 1.56 & 1.62 & 3.30 \\\cline{2-7}
 $3^+$ &$^{34}$Si & 4.74 & 5.70 & 4.00 & 4.91 & ---\\
   &  $^{36}$S & 7.68 & 8.20 & 7.63 & 8.10 & 5.46  \\\cline{2-7}
 $4^+$ &$^{34}$Si & 6.77 & 6.78 & 6.64 & 6.42 & ---\\ 
   &  $^{36}$S & 7.15 & 7.12 & 7.09 & 6.83 & 6.51 \\\cline{2-7}
 $1^+$ &$^{34}$Si & 8.12 & 9.12 & 5.00 & 7.46 & --- \\
  & $^{36}$S &8.47 & 9.37 & 8.40 & 8.38 & 4.52 \\
\hline 
 $2^+$ &$^{34}$Ca &  3.55 & 3.88 & 3.86 & 3.85 & ---\\  
  &$^{36}$Ca & 2.51 & 2.72 & 2.39 & 2.49 &--- \\ \cline{2-7}
 $3^+$ &$^{34}$Ca &  4.68 & 5.58 & 4.70 & 4.70 & ---\\    
  &$^{36}$Ca & 8.14 & 8.48 & 8.51 & 8.59 & ---\\ \cline{2-7}
 $4^+$ &$^{34}$Ca &  6.73 & 6.73 & 6.72 & 6.31 & --- \\
 &$^{36}$Ca & 7.55 & 7.31 & 7.33 & 7.13 & ---\\    \cline{2-7}
 $1^+$ &$^{34}$Ca &  8.15 & 9.13 & 7.72 & 7.44  & ---\\
  &$^{36}$Ca & 9.01 & 9.78 & 8.74 & 8.77 &--- \\ \cline{2-7}
 \hline \hline
\end{tabular}
\vskip -0.2 cm
\caption{
Excitation energies, expressed in MeV, of $^{34}$Si, $^{36}$S, $^{34}$Ca 
and $^{36}$Ca nuclei. The experimental data are from Ref. \cite{bnlw}.
}
\label{tab:spectrum} 
\end{center} 
\end{table} 

We investigated
excited states dominated by particle-hole configurations 
involving s.o. partner levels. This implies the study of positive 
parity states, $0^+$ excitations excluded. We have carried 
out our calculations by using the QRPA approach 
described in detail in Ref. \cite{don17}.  

We present, in Table \ref{tab:spectrum}, the results obtained 
for the $N=20$ isotones $^{34}$Si and $^{36}$S and  
for the $Z=20$ isotopes $^{34}$Ca and $^{36}$Ca.
In these nuclei the shell closure at 20 induces 
low-lying excitations dominated by the protons in
$^{34}$Si and $^{36}$S and by the neutrons in
the calcium isotopes. We have selected these nuclei 
since the open shells are filled by 14 or 16 
nucleons where we found the main effect of the 
SB structure on the s.p. energies.

In Table \ref{tab:spectrum} we show the excitation 
energies for the positive parity states obtained
with the four interactions we have considered, and we compare 
them with available experimental data, taken from 
the compilation of Ref. \cite{bnlw}.
We observe that the differences between the
energies obtained with the four interactions are 
1.5 MeV at most, with the exception of the 
$1^+$ state in $^{34}$Si that shows a rather small excitation 
energy in the case of the D1MTd force. In this case, 
the tensor terms generate high collectivity 
in the wave function that is not present 
when the other interactions are considered.

Remarkable discrepancies with the available
experimental data are those of the 
$2^+$ state in $^{36}$S, which is underestimated by 
more than $1.5\,$MeV, and the $1^+$ and $3^+$ states 
in the same nucleus, overestimated by about $4\,$MeV and $2\,$MeV, respectively.
In the $4^+$ case the differences between our results
and the experimental data are about $1\,$MeV.

When going from $^{34}$Si to $^{36}$S and from 
$^{34}$Ca to $^{36}$Ca, 
the energy of the $2^+$ state diminishes of
about $2\,$MeV and $1.5\,$MeV, respectively. 
For the other excited states analyzed, the energies increase, 
specially for the $3^+$ states. We have to remark, however, 
that the $3^+$ states 
in the nuclei with $A=34$ are dominated by the $(2s_{1/2}, 1d_{5/2})$ 
configuration while for $A=36$ the main configuration is $(1d_{3/2},1d_{5/2})$. 
As a consequence a direct relation between the change in the s.o. 
splitting and the variation in the excitation energy cannot be established. A similar 
situation occurs for the $2^+$ excited states.

\begin{table}[!t] 
\begin{center} 
\begin{tabular}{ccccccccccccccc}
\hline \hline
\multicolumn{15}{c}{$4^+$} \\ \hline
& $\omega$ & \multicolumn{2}{c}{$(1d_{3/2}, 1d_{5/2})_{\rm p}$} &~~& \multicolumn{2}{c}{$(1d_{3/2}, 1d_{5/2})_{\rm n}$} &
 & & $\omega$ & \multicolumn{2}{c}{$(1d_{3/2}, 1d_{5/2})_{\rm p}$} &~~& \multicolumn{2}{c}{$(1d_{3/2}, 1d_{5/2})_{\rm n}$} \\
 \cline{3-4} \cline{6-7}  \cline{11-12}  \cline{14-15}
 nucleus & (MeV) & $|X|$ & $|Y|$ &~~& $|X|$ & $|Y|$ & ~~& 
            nucleus & (MeV) & $|X|$ &$|Y|$ &~~& $|X|$ & $|Y|$ \\
\hline
 $^{30}$Si &  6.32 & 0.74 & 0.09 && 0.67 & 0.09 & & 
 $^{32}$S &  5.90 & 0.73 & 0.10 &&   0.68 & 0.10 \\ 
 $^{32}$Si &  6.54 & 0.84 &  0.07 && 0.52 & 0.05 & &
 $^{34}$S &  6.41 & 0.82 &  0.07 && 0.55 & 0.06 \\ 
 $^{34}$Si &  6.77 & 0.99 & 0.05 & & --- & --- & &
  $^{36}$S &  7.15 & 0.99 & 0.03 & & --- & --- \\ \hline
  
 $^{30}$S &  6.96 & 0.74 & 0.08 && 0.66 & 0.08 & & 
 $^{32}$S &  5.90 & 0.73 & 0.10 &&   0.68 & 0.10 \\ 
 $^{34}$Ca &  6.73 & --- & --- & & 0.99 & 0.05 & &
  $^{36}$Ca &  7.55 &  --- & --- && 0.99 & 0.03 \\ 
 \hline\hline
\end{tabular}
\vskip -0.2 cm
\caption{
Excitation energies of the $4^+$ states in various silicon, sulfur, argon and calcium isotopes 
and QRPA amplitudes of the main configurations of the corresponding 
wave functions, obtained by using the D1M interaction. 
}
\label{tab:fourp} 
\end{center} 
\end{table} 

\begin{table}[!b] 
\begin{center} 
\begin{tabular}{ccccc} 
\hline \hline
 & D1M & D1S & D1MTd & D1ST2a  \\ \hline
$s^{\rm p}_{1d}(^{36}{\rm S}) - s^{\rm p}_{1d}(^{34}{\rm Si})$ (MeV)  & 0.62 & 1.41 & 0.50 & 0.51  \\
$\omega_{4^+}(^{36}{\rm S}) - \omega_{4^+}(^{34}{\rm Si})$ (MeV)   & 0.38 & 0.34 & 0.45 & 0.41 \\
                         $\fr_{\rm p}(^{36}{\rm S}) - \fr_{\rm p}(^{34}{\rm Si})$ & $-0.24$ & $-0.49$ & $-0.30$ & $-0.51$ \\  
                        $\dr_{\rm p}(^{36}{\rm S}) - \dr_{\rm p}(^{34}{\rm Si})$ & $-0.24$ & $-0.36$ & $-0.24$ & $-0.38$ \\  
\hline
$s^{\rm n}_{1d}(^{36}{\rm Ca}) - s^{\rm n}_{1d}(^{34}{\rm Ca})$ (MeV)  & 1.10 & 1.02 & 0.78 & 1.06  \\
$\omega_{4^+}(^{36}{\rm Ca}) - \omega_{4^+}(^{34}{\rm Ca})$ (MeV)   & 0.81 & 0.59 & 0.62 & 0.82 \\
                         $\fr_{\rm n}(^{36}{\rm Ca}) - \fr_{\rm n}(^{34}{\rm Ca})$ & $-0.17$ & $-0.31$ & $-0.23$ & $-0.29$ \\  
                        $\dr_{\rm n}(^{36}{\rm Ca}) - \dr_{\rm n}(^{34}{\rm Ca})$ & $-0.13$ & $-0.23$ & $-0.17$ & $-0.23$ \\  
 \hline \hline
\end{tabular}
\vskip -0.2 cm
\caption{Differences between various quantities calculated in the two 
pairs of nuclei $^{36}$S$-^{34}$Si and $^{36}$Ca$-^{34}$Ca. Specifically, 
the differences between (i) the s.p. energies of the proton or neutron $d$ levels; (ii) the QRPA energies of  the $4^+$ excitations; (iii)
the values of the depletion fraction $\fr$, defined in  Eq.~(\ref{eq:fr}), and (iv) those of the flatness index $\dr$, defined in Eq.~(\ref{eq:dr}) are given..
}
\label{tab:speom} 
\end{center} 
\end{table} 

To avoid this problem we have focussed our attention 
on $4^+$ excitations that are rather 
well selective for the configurations involving the $1d$ s.o. 
partner levels. We summarize in
Table \ref{tab:fourp} the results obtained for various 
isotopes grouped in pairs that have the same number of either 
neutrons or protons and, simultaneously have either $Z=14,16$ 
or $N=14,16$. We include only the results obtained with the 
D1M interaction, those found for the other forces being similar. 
In all the cases, the dominant p-h configurations are the 
proton $(1d_{3/2}, 1d_{5/2})_{\rm p}$ and the neutron 
$(1d_{3/2}, 1d_{5/2})_{\rm n}$. Specifically, we show
the energies of the $4^+$ excited states and the absolute 
values of the QRPA amplitudes $X$ and $Y$ \cite{don17} when $|X|>0.1$. 

We observe that only in the $^{34}$Si and $^{36}$S nuclei the excitation
is dominated by a single proton $(1d_{3/2}, 1d_{5/2})_{\rm p}$ configuration. 
In a similar way, only in the case of the $^{34}$Ca and $^{36}$Ca 
the excited state is an almost pure neutron $(1d_{3/2}, 1d_{5/2})_{\rm n}$ configuration. 
In the other
nuclei, the opening of both neutron and proton shells allows a mixture
of the $(1d_{3/2}, 1d_{5/2})_{\rm p}$ and $(1d_{3/2}, 1d_{5/2})_{\rm n}$ 
components in the wave functions. 
In these latter nuclei, the excitation energy is a kind of average 
of the energies of these two configurations. 

\begin{figure}[!ht] 
\begin{center} 
\includegraphics [scale=0.5,angle=0]{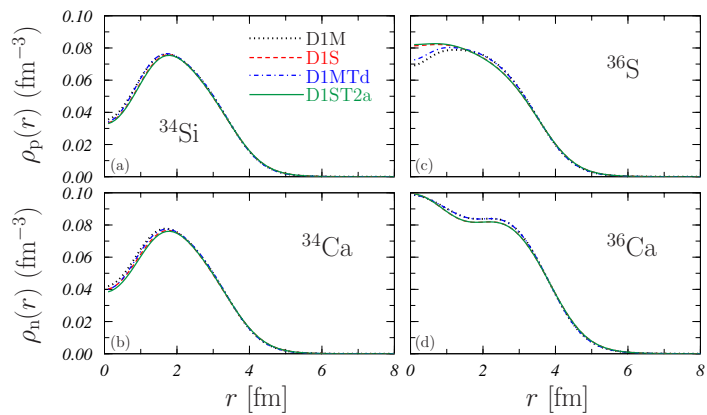} 
\vskip 0.0 cm 
\caption{\small 
Density distributions of protons, panels (a) and (c), 
and neutrons, panels (b) and (d), of the nuclei analyzed 
in Table \ref{tab:speom}. The four lines indicate the results obtained by 
using the four interactions considered in the present work. 
 }
\label{fig:dens2} 
\end{center} 
\end{figure}

In Table \ref{tab:speom} we compare, for the four interactions, 
the results obtained for various quantities
related to the SB structure of the density.
For the two pairs of nuclei ($^{36}$S$-^{34}$Si and
 $^{36}$Ca$-^{34}$Ca), we have evaluated the differences
between the respective s.o. splittings $s_{1d}^\alpha$, 
excitation energies of the $4^+$
states, $\omega_{4^+}$, depletion fraction $\fr_\alpha$ 
and flatness index $\dr_\alpha$. As we can see, 
a reduction in $\fr$, and in $\dr$, 
occurs when going from $A=34$ to $A=36$, indicating that 
the corresponding density looses its SB structure, as it
is clearly shown  in  Fig. \ref{fig:dens2}.
This is related to an increase of the excitation energy of the $4^+$ 
state due to an enhancement in the s.o. splitting of the $1d$ 
s.p. level. This happens for the two pairs of nuclei and for all the interactions.

From what we have discussed the identification of $4^+$ states in the
spectra of the four nuclei considered can be used to infer the
presence of a SB structure in the $A=34$ nuclei. As it is shown in
Table \ref{tab:spectrum} only the $4^+$ state in $^{36}$S has been
identified at about $6.51\,$MeV. By considering the range of uncertainty
of our calculations, related to the use of different nucleon-nucleon
interactions, we would expect a $4^+$ state in $^{34}$Si between 6.0 and
$6.2\,$MeV. The state identified at $6.023\,$MeV 
and whose multipolarity has not yet been assigned 
\cite{bnlw} could be that $4^+$ level. 

%%%%%%%%%%%%%%%%%%%%%%%%%%%%%%%%%%%%
\section{Conclusions}
\label{sec:conclusions}
In this article we have investigated the possibility that some nuclei
present a SB structure in their proton, neutron or matter
distributions. Since a direct identification of these structures
requires very involved scattering experiments difficult to be carry 
out on unstable targets, we have explored the possibility that the SB
distributions can be linked to observables more easily measurables. 
The
relationship between the derivative of the distributions and the
s.o. interaction induced us to study the energy
splittings of s.o. partner levels, and also the
energy spectra.

In our investigation we used a HF+BCS approach that generates the
s.p. bases required as input for our QRPA calculations of the excited
states. These calculations have been done for spherical nuclei, 
by using the same interaction in all the three steps, HF,
BCS and QRPA. We used four different parameterizations of 
the finite-range
Gogny force, two of them containing tensor terms. 
In this way we have estimated the sensitivity 
of our results to the only physical input of our approach: the
effective nucleon-nucleon interaction.

The validity of our calculations has been tested against the
experimental binding energies, and we found excellent agreements.
Also the experimental values of the charge rms radii are rather well
reproduced. We have compared the charge distributions obtained in our
calculations with the available empirical ones. While their
behaviors are well reproduced on the 
nuclear surface, there are some discrepancies in the interior, where the
SB structures appear. 

However, since investigating the nuclear interior is rather
difficult we
have studied the effects of the SB structures on 
other observables such as those linked to the
s.o. interaction. We have pragmatically verified that there is a
relationship between the occurrence of SB 
structures in the density distributions and the size of the splitting
between the energies of s.p. levels that are s.o. partners.  
The general trend we found is that the s.o. 
splittings in isotones, or isotopes, 
with 14 protons, or neutrons, that have SB structures in their
proton, or neutron densities, 
 are smaller than those with 16, where SB distributions do not occur. 
The differences are relatively large and this behaviour occurs 
in all the calculations we have carried
out, independently of the interaction used and 
all along the various isotope and isotone
chains studied. 

This modification of the s.o. energy splitting has consequences
on the excitation spectrum. We have studied with our QRPA theory the
positive parity excited states of various nuclei. 
We have found that the most interesting cases are the 
low-lying $4^+$ states in the isotones
$^{34}$Si and $^{36}$S, and in the isotopes
$^{34}$Ca and $^{36}$Ca. 
In these nuclei, the $4^+$ excitation is dominated by a single, 
almost pure configuration formed by the $1d$ s.o. partners levels. 
We have found that the   
$4^+$ excited states in $A=34$ nuclei have lower energies than the
analogous ones in the $A=36$ nuclei.

At present only a $4^+$ state in $^{36}$S at $6.5\,$MeV is
known. Our calculations predict a $4^+$ state in $^{34}$Si 
at about 6.0 MeV. The identification of this state would validate our
approach and indicate the existence of a SB structure in $^{34}$Si.

%%%%%%%%%%%%%%%%%%%%%%%%%%%%%%%%%%%%% 
\acknowledgments  
This work has been partially supported by  
the Junta de Andaluc\'{\i}a (FQM387), the Spanish Ministerio de 
Econom\'{\i}a y Competitividad (FPA2015-67694-P) and the European 
Regional Development Fund (ERDF). 
 
%%%%%%%%%%%%%%%%%%%%%%%%%%%%%%%%%%%%%%%%%%%%%%%%%%%%%% 
%   Bibliography 
%%%%%%%%%%%%%%%%%%%%%%%%%%%%%%%%%%%%%%%%%%%%%%%%%%%%%% 

\end{document}